\documentstyle[prl,aps,multicol,epsf]{revtex}
\begin{document}
\draft
\title{Magnetoconductivity of Insulating Silicon Inversion Layers}

\author{Yeekin Tsui, S. A. Vitkalov and M.P. Sarachik}
\address{Physics Department, City College of the City
University of New York, New York, New York 10031}
\author{T.~M.~Klapwijk}
\address{Delft University of Technology, Department of Applied Physics,
2628 CJ Delft, The Netherlands}
\date{\today}

\maketitle

\begin{abstract}

The normalized in-plane magnetoconductivity of the dilute strongly
interacting system of electrons in silicon MOSFET's scales with $B/T$ for low
densities in the insulating phase. Pronounced deviations occur at higher
metallic-like densities, where a new energy scale $k_B \Delta$ emerges which
is not associated with either magnetic field or thermal effects.  $B/T$
scaling of the magnetoconductivity breaks down at the density $n_0$ where the
energy scale $k_B \Delta$ vanishes, near or at the critical density
$n_c$ for the apparent metal-insulator transition.  The different behavior
of the magnetoconductivity at low and high densities suggests the existence of
two distinct phases.

\end{abstract}


\begin{multicols}{2}

According to well established theory, no metallic state can exist in
two dimensions for noninteracting \cite{Abrahams79} or weakly interacting
\cite{aa} electrons (or holes) in zero magnetic field in the limit
of zero temperature.  In dilute two-dimensional systems where the
interactions are known to be strong, however, experimental studies have
revealed an unexpected decrease in the resistance as the temperature is
lowered, behavior that is generally a characteristic of metals.  This
metallic behavior has been observed down to the lowest accessible
temperatures at electron \cite{Krav94,PoFoWa97} and hole
\cite{Cole97,Hanein98,Simmons98} densities above some critical density
$n_c$ (or $p_c$).  For densities down to approximately $1.5 n_c$, the
temperature dependence has been attributed to electron-electron
scattering in the ballistic limit ($k_BT>>\hbar/\tau)$, confirming the
importance of strong e-e interactions in determining the behavior of these
systems \cite{zala}.  However, the behavior observed at lower densities and
the nature of the apparent metal-insulator transition are still not understood
\cite{AbKrSa01}.

A very unusual characteristic of dilute, strongly interacting electron
(hole) systems is their strong response to an in-plane magnetic field: the
resistivity increases dramatically with increasing field and saturates
to a new value above a characteristic magnetic field that depends on
density and temperature \cite{AbKrSa01}.  Simonian {\it et al.} reported
that for small to moderate magnetic fields, the  magnetoconductivity of
silicon MOSFET's scales as $(B/T)$ for electron densities at and slightly
above $n_c$ \cite{Simonian98}.  From an analysis of the temperature and
density dependence of the magnetoconductance of silicon MOSFET's, Vitkalov
{\it et al.} \cite{Vitkalov01} have identified an energy scale $(k_B
\Delta)$ which extrapolates to zero at a finite density $n_0$ in the
vicinity of $n_c$; the behavior of the magnetoconductivity was attributed
to an increase in the magnetic susceptibility $\chi \propto$ (g$^*$ m$^*)$
and the approach to a zero temperature quantum phase transition at $n_0$
(here g$^*$ and m$^*$ are the renormalized Lande-g factor and effective
mass, respectively).  Shashkin {\it et al.} \cite{Shashkin01} found
similar results; moreover, based on measurements of Shubnikov-de Haas
oscillations and in-plane magnetoresistivity, these authors recently
\cite{Shashkin02} reported that the sharp increase in the susceptibility is
associated with an increase in the effective mass while the $g$-value remains
essentially constant as the electron density approaches $n_c$.  These
findings all suggest critical behavior and the approach to a quantum phase
transition.

In this paper we report the results of an investigation of the
magnetoconductivity of the dilute two-dimensional system of electrons in
a high-mobility silicon MOSFET in the insulating phase.  We find that the
in-plane magnetoconductivity scales with $(B/T)$ for electron densities
near and below $n_0$.  Pronounced deviations from this simple
scaling form become evident in the metallic regime at higher densities: in
agreement with our earlier findings \cite{Vitkalov01}, scaling in the
metallic phase requires the inclusion of an additional energy scale
$(k_B\Delta)$.  The different behavior
of the magnetoconductivity at low and high densities suggests the existence of
two distinct phases.

The sample used in these studies is a high-mobility silicon MOSFET
($\mu_{\textrm{peak}} \sim 20000/Vs$ at $0.5$ K).  Contact resistances
were minimized by using a split-gate geometry that allows a higher
electron density to be established in the vicinity of the contacts than in
the 2D system under investigation.  Data were taken by standard
four-terminal ac techniques for electron densities above $1.2 \times
10^{11}~\textrm{cm}^{-2}$, and by dc techniques for lower densities.
Experiments were performed at temperatures between $0.25$ K and $4$ K in
magnetic fields up to $10$ T; data were taken in the linear regime
using small currents to avoid overheating the electrons.  The critical
density of the sample is $\approx 0.90 \times 10^{11}$ cm$^{-2}$.

\vbox{
\vspace{0.4 in}
\hbox{
\hspace{-0.2in} 
\epsfxsize 3.4 in \epsfbox{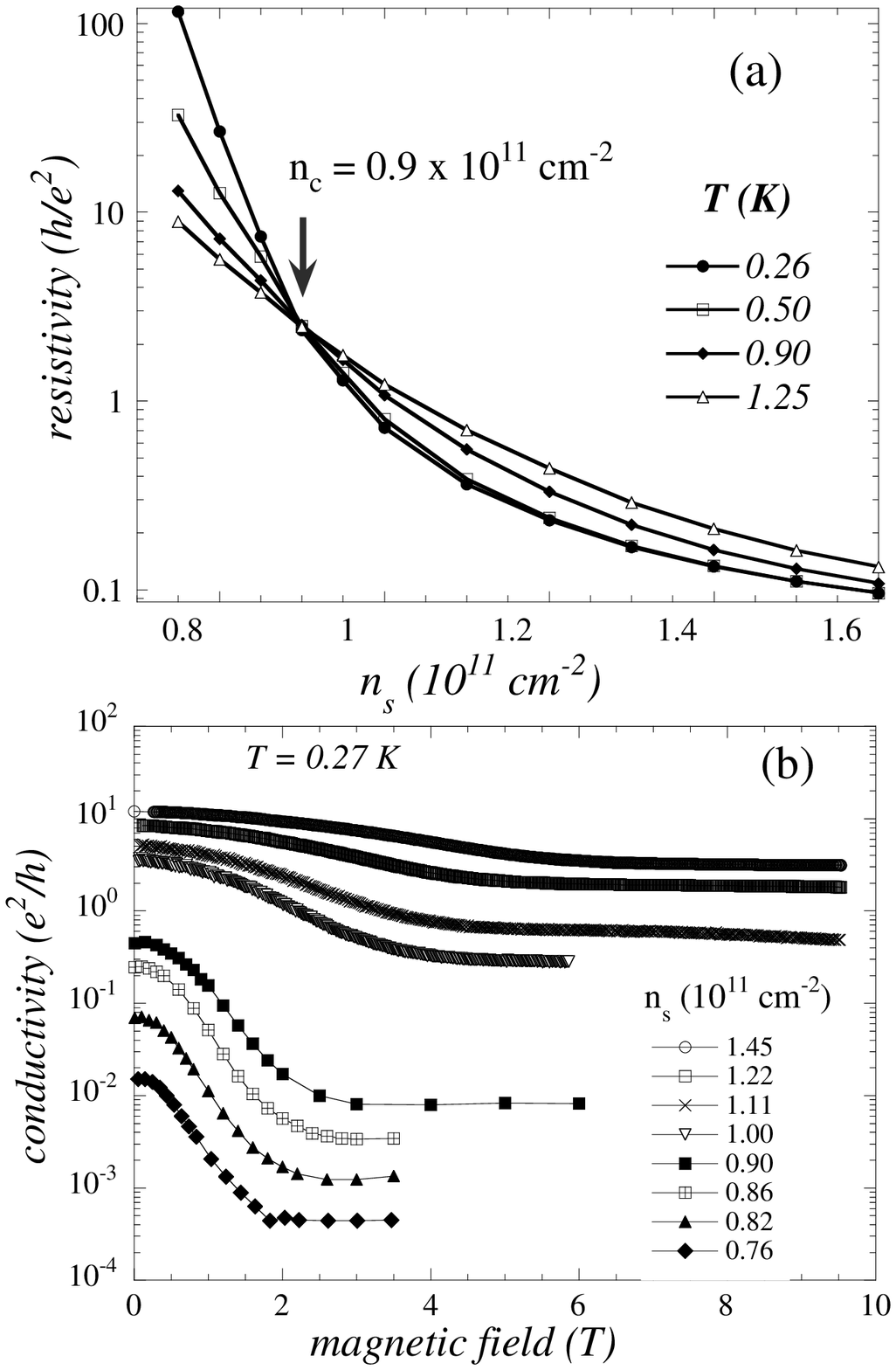} 
}
}
\vskip 0.5cm
\refstepcounter{figure}
\parbox[b]{3.1in}{\baselineskip=12pt FIG.~\thefigure.
(a) Resistivity of a silicon MOSFET at four temperatures,
as labeled, as a function of electron density.  The curves cross at
the critical density $n_c = 0.9 \times 10^{11}$ cm$^{-2}$.  (b)
Conductivity as a function of applied in-plane magnetic field at
different densities, as labeled; $T = \sim$ 0.25 K.  The
top four magnetoconductance curves are at metallic densities, and the
bottom three are insulating.

\vspace{0.10in}
}
\label{raw} 

Figure \ref{raw}(a) shows the resistivity of the silicon MOSFET as a
function of electron density at four different temperatures: the behavior
is insulating for densities below the crossing point at $0.9 \times
10^{11}$ cm$^{-2}$ (resistivity increasing with decreasing temperature)
and metallic above that density.  Figure \ref{raw}(b) shows the
conductivity as a function of in-plane magnetic field for different
densities at $T \sim 0.25$ K; here the top four curves correspond to
densities in the conducting range, the fifth is at or near the critical
density, and the remaining are in the insulating regime.  Consistent with
earlier results in silicon and various materials, the magnetocoductivity
saturates at a progressively lower applied field as the density
decreases.  The enormous response to in-plane magnetic field is a
typical feature in these 2D systems
\cite{Dolg92,Simonian97,Pudalov97,Yoon99}.

\vbox{
\vspace{0.4 in}
\hbox{
\hspace{-0.2in} 
\epsfxsize 3.4 in \epsfbox{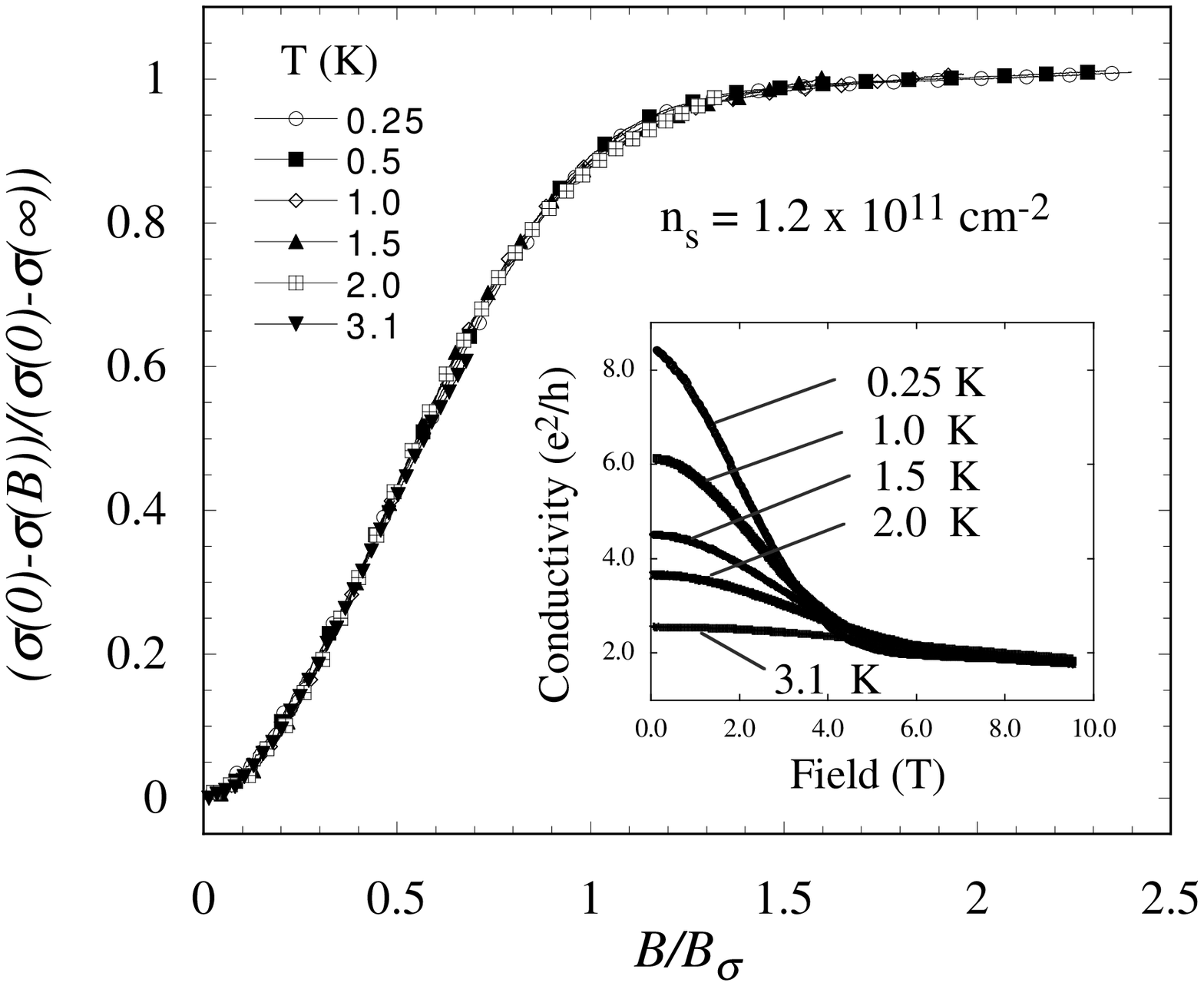} 
}
}
\vskip 0.5cm
\refstepcounter{figure}
\parbox[b]{3.1in}{\baselineskip=12pt FIG.~\thefigure.
Normalized magnetoconductivity as a function of the scaled field
at different temperatures at electron density $n_s = 1.2 \times 10^{11}~
\textrm{cm}^{-2}$. The inset shows the (unnormalized) conductivity versus
in-plane magnetic field at various temperatures.

\vspace{0.10in}
}
\label{metallic}

For a fixed electron density in the metallic regime ($n_s>n_c)$, the inset
to Fig. \ref{metallic} shows the conductivity as a function of
in-plane magnetic field at different temperatures.  Following the
procedure used in a previous study \cite{Vitkalov01}, the normalized
magnetoconductivity:
\begin{equation} \label{eqn:theratio}
\sigma_{\textrm{norm}} \equiv \frac{\sigma(B=0) -
\sigma(B)}{\sigma{(B=0)} -
\sigma{(B\rightarrow\infty)}}~,
\end{equation}
is plotted as a function of a scaled magnetic field $B/B_\sigma$.  Here
$B_\sigma$ is a fitting (or scaling) parameter chosen to obtain a data
collapse.  Note that the normalized magnetoconductivity
(\ref{eqn:theratio}) is simply the field dependent contribution to the
conductivity,
$[\sigma(B=0) - \sigma(B)]$, normalized by its full value, $[\sigma(B=0)
- \sigma(B\rightarrow\infty)]$.

A study of the scaled magnetoconductivity at various temperatures and
densities allows a determination of the parameter $B_\sigma$ as a
function of temperature and density.  Consistent with our earlier
analysis \cite{Vitkalov01},
$B_\sigma$ satisfies the empirical formula:
\begin{equation} \label{eqn:Bsigma}
B_\sigma = A_n\sqrt{\Delta^2 + T^2}~,
\end{equation}
where the fitting parameter $A_n$ varies by less than 15\% in the range
of densities of our experiments.  The result of this analysis is
shown in Fig. 3, where $\Delta$ is plotted as a function of electron
density
$n_s$.

\vbox{
\vspace{0.4 in}
\hbox{
\hspace{-0.2in} 
\epsfxsize 3.4 in \epsfbox{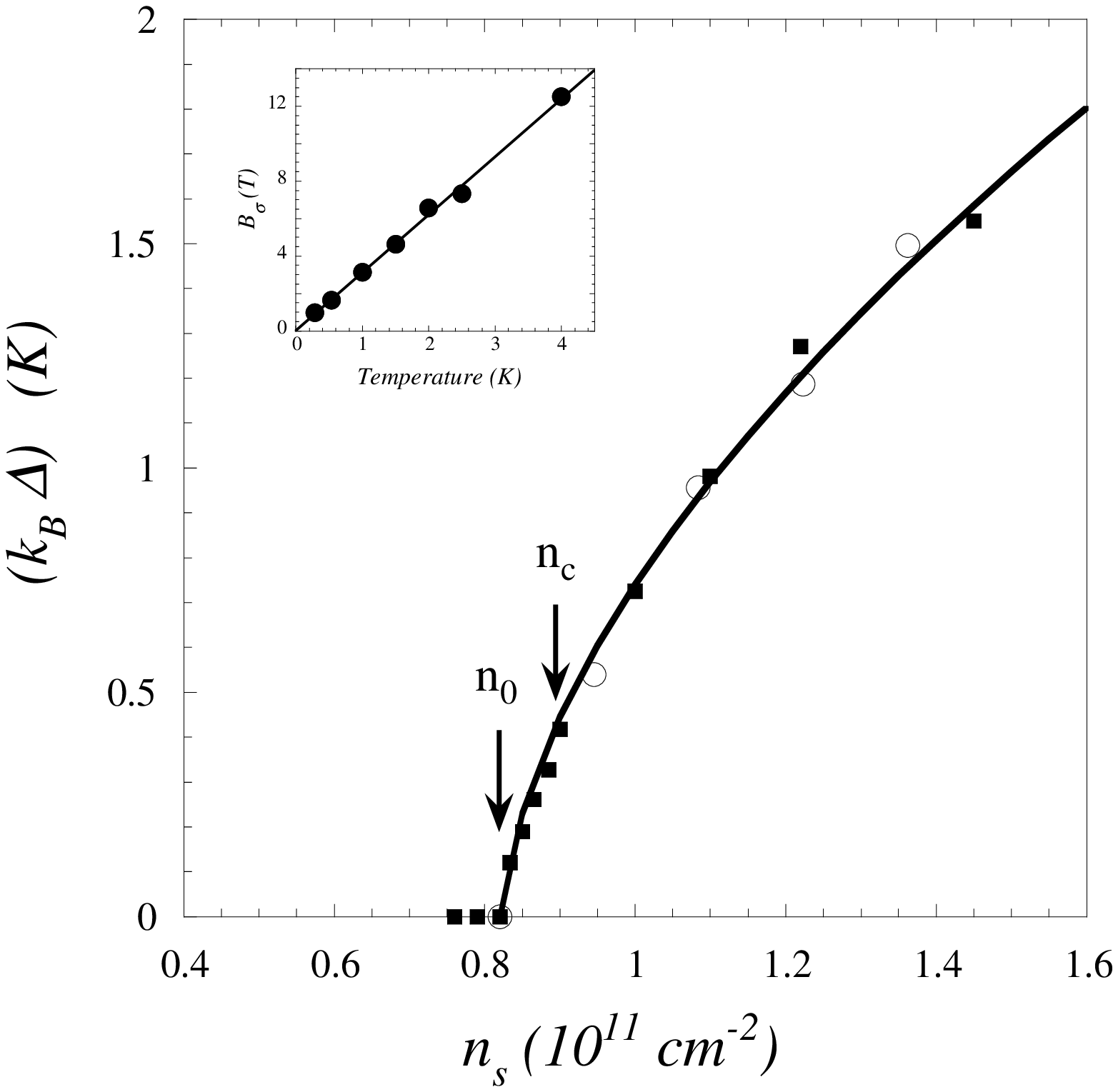} 
}
}
\vskip 0.5cm
\refstepcounter{figure}
\parbox[b]{3.1in}{\baselineskip=12pt FIG.~\thefigure.
The parameter $\Delta$ determined from measured data using Eq.
\ref{eqn:Bsigma} versus electron density $n_s$.  The open circles
represent data of Vitkalov {\it et al.} \cite{Vitkalov01} and closed symbols
denote new data reported in this paper; the line is a guide to the eye.  The
inset shows
$B_\sigma$ versus temperature $T$ for density $n_s = 0.82 \times 10^{11}~
\textrm{cm}^{-2}$, illustrating that
$B_\sigma = A_nT$ and $\Delta = 0$ for this density.

\vspace{0.10in}
}
\label{delta}

The quantity $\Delta$ enters on an equal footing with the temperature
$T$ and represents an energy scale $(k_B \Delta)$ associated with
$B_\sigma$.  In agreement with published results, the plot shown in Fig.
\ref{delta} indicates that the energy $(k_B \Delta)$ extrapolates to
zero at a finite electron density $n_0$ close to the critical density
$n_c$ that signals the change in temperature dependence of the
conductivity from metallic to insulating \cite{Vitkalov01}.  That $\Delta
= 0$ at $n_0$, and hence, $B_\sigma = A_nT$, is illustrated in the inset
to Fig.
\ref{delta}, where we have used $B_\sigma$ as a free fitting
parameter to collapse the curves; the parameter $\Delta$ equals zero with
an estimated accuracy of 5-10 mK.

The vanishing energy $(k_B\Delta)$ signals a diverging correlation time
scale $\tau \sim\hbar/(k_B\Delta)$, suggesting that the system is
approaching a quantum phase transition in the limit $T
\rightarrow 0$.

The results of our current measurements are denoted by black squares
in Fig. 3.   For the metallic densities above $n_c$, the new measurements
provide additional and more precise data that support our earlier
conclusions.  At low densities, our measurements
provide important new information: the energy scale $(k_B\Delta)$ remains
zero in the insulator down to the lowest density measured, $n=0.76
\times 10^{11}$ cm$^{-2}$, $\sim 15$ \% below the critical density.  This
implies that the normalized magnetoconductivity scales strictly with $B/T$
in the insulating phase (see Eq. \ref{eqn:Bsigma}).

\vbox{
\vspace{0.4 in}
\hbox{
\hspace{-0.2in} 
\epsfxsize 3.4 in \epsfbox{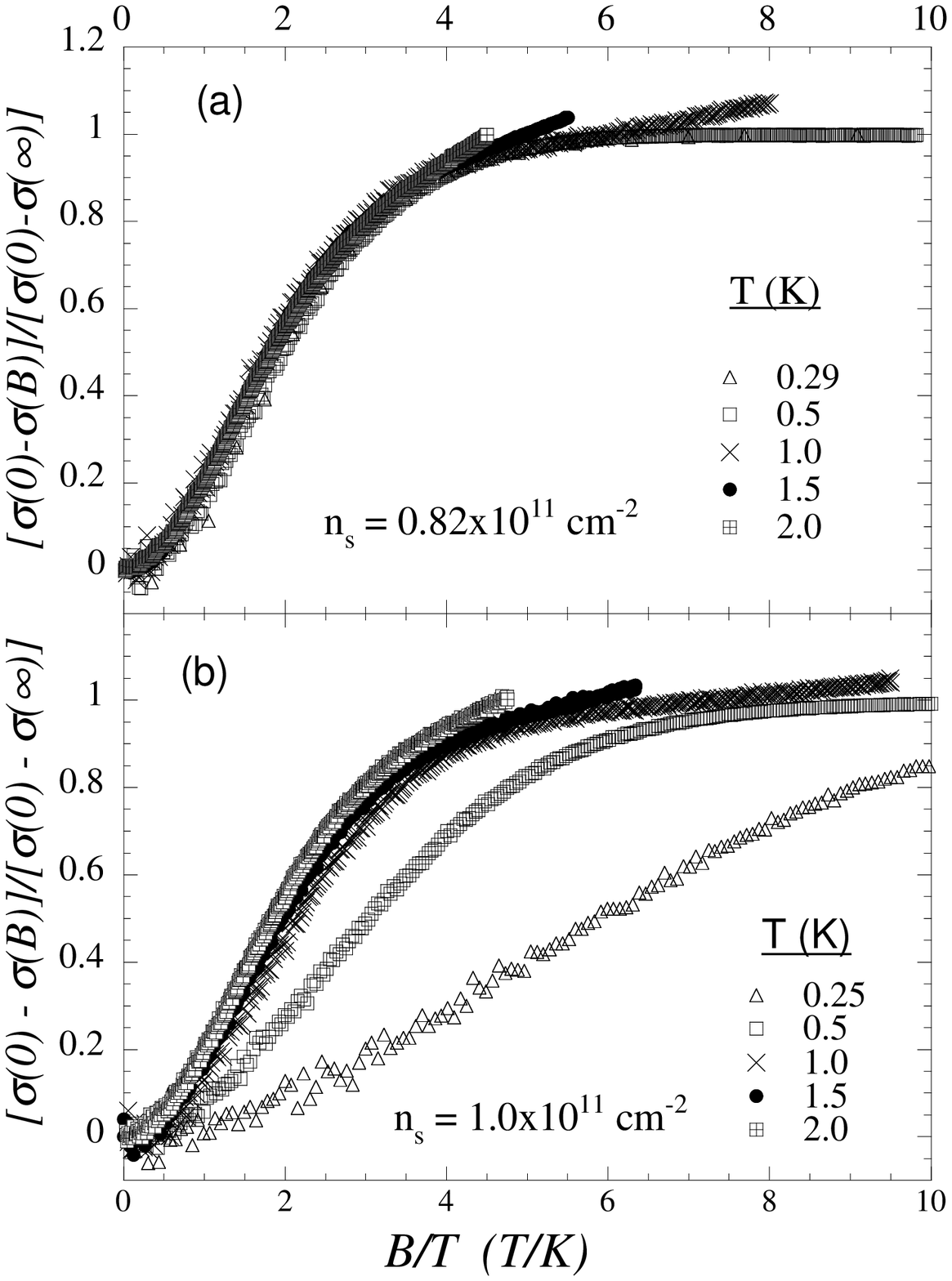} 
}
}
\vskip 0.5cm
\refstepcounter{figure}
\parbox[b]{3.1in}{\baselineskip=12pt FIG.~\thefigure.
The normalized magnetoconductivity (see Eq. \ref{eqn:theratio})
versus $B/T$ for (a) insulating density $n_s = 0.82 \times 10^{11}$
cm$^{-2}$, and (b) metallic density $n = 1.0 \times 10^{11}$ cm$^{-2}$.

\vspace{0.10in}
}
\label{temps}

This is shown explicitly in Figs. \ref{temps} and
\ref{densities}.  Fig. \ref{temps} shows the normalized
magnetoconductivity as a function of
$B/T$ at various different temperatures for two electron densities, one below
and one above $n_c$: the magnetoconductivity curves collapse onto one curve
in the insulator shown in frame (a) while this is clearly not true for
the metallic density shown in frame (b).  Fig. \ref{densities} shows the
normalized magnetoconductivity as a function of $B/T$ at temperature $0.5$
K for various densities across the transition.  The data collapse onto a
single curve for densities up to $0.9 \times 10^{11}$ cm$^{-2}$, and
deviations become progressively more pronounced as the electron density
is increased into the metallic phase.

Experiments on MOSFET's \cite{okamoto,vitkalovSdH} and other two-dimensional
materials indicate that the magnetoconductivity is directly related to the
degree of polarization of the electrons, (holes), and therefore, to the
magnetization.  This one-to-one correspondence suggests that if the
conductivity scales with $B/T$, then so does the magnetization.  We note that
the magnetization of a set of localized, noninteracting magnetic moments is a
function of $B/T$.  However, this is quite unlikely to provide an adequate
description for dilute 2D silicon MOSFET's, where electron interactions are
an order of magnitude larger than the Fermi energy and provide the dominant
energy in the problem. 
$B/T$ behavior is predicted in the insulating phase by Agrinskaya and Kozub,
who calculated the effect of on-site spin correlations on the hopping
conductivity \cite{AgKo98}.

\vbox{
\vspace{0.4 in}
\hbox{
\hspace{-0.2in} 
\epsfxsize 3.4 in \epsfbox{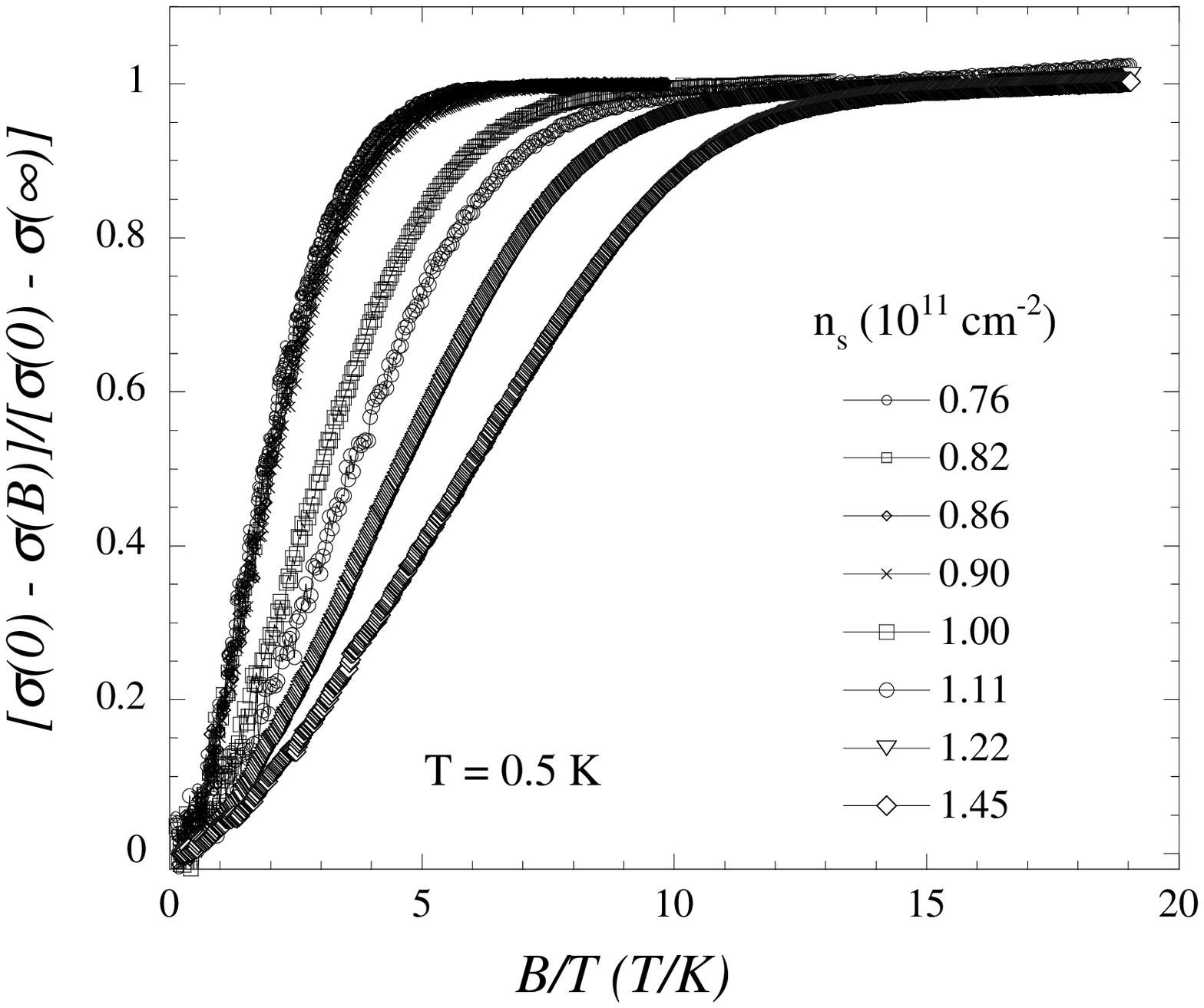} 
}
}
\vskip 0.5cm
\refstepcounter{figure}
\parbox[b]{3.1in}{\baselineskip=12pt FIG.~\thefigure.
The normalized magnetoconductivity (see Eq. \ref{eqn:theratio})
versus $B/T$ for several electron densities at $T = 0.5$ K.  Note that the
magnetoconductivity scales with $B/T$ for insulating densities $n_s < n_c$
and does not scale for metallic densities.
\vspace{0.10in}
}
\label{densities}

To summarize, the normalized in-plane magnetoconductivity of the dilute
strongly interacting system of electrons in silicon MOSFET's scales with
$B/T$ in the insulating regime for densities below $n_0$.  Deviations
occur at higher metallic-like densities, where a new energy scale emerges
which is not associated with either magnetic field or thermal effects.  The
breakdown of $B/T$ scaling of the magnetoconductivity above $n_0$, which is at
or near the electron density $n_c$ that signals the change from insulating to
metallic temperature dependence of the conductivity, suggests that
distinct phases exist above and below $n_0$.  Our finding of
$B/T$ behavior in the insulating regime sets an important constraint on
theory.

We thank B. Spivak for useful discussions.  This work was supported by
DOE-FG02-84-ER45153 and NSF grant DMR-0129581.

\end{multicols}


\begin{thebibliography}{99}
\bibitem{Abrahams79}
E. Abrahams, P. W. Anderson, D. C. Licciardello, and T. V. Ramakrishnan,
Phys. Rev. Lett. {\bf 42}, 673 (1979).
\bibitem{aa}
B. L. Altshuler, A. G. Aronov, and P. A. Lee, Phys. Rev. Lett., {\bf 44},
1288 (1980).
\bibitem{Krav94}
S. V. Kravchenko, G. V. Kravchenko, J. E. Furneaux, V. M. Pudalov, and M.
D'Iorio, Phys. Rev. B {\bf 50}, 8039 (1994);
S. V. Kravchenko, W. E. Mason, G. E. Bowker, J. E. Furneaux, V. M.
Pudalov, and M. D'Iorio, Phys. Rev. B {\bf 51}, 7038 (1995);
S. V. Kravchenko, D. Simonian, M. P. Sarachik, Whitney Mason, and J. E.
Furneaux, Phys. Rev. Lett. {\bf 77}, 4938 (1996); S. V. Kravchenko and T. M.
Klapwijk, Phys. Rev. Lett. {\bf 84} 2909 (2000).
\bibitem{PoFoWa97}
D. Popovi\'{c}, A. B. Fowler, and S. Washburn, Phys. Rev. Lett. {\bf 79},
1543 (1997).
\bibitem{Cole97}
P. T. Coleridge, R. L. Williams, Y. Feng, and P. Zawadzki, Phys. Rev. B
{\bf 56}, R12764 (1997).
\bibitem{Hanein98}
Y. Hanein, U. Meirav, D. Shahar, C. C. Li, D. C. Tsui, and H. Shtrikman,
Phys. Rev. Lett. {\bf 80}, 1288 (1998).
\bibitem{Simmons98}
M. Y. Simmons, A. R. Hamilton, M. Pepper, E. H. Linfield, P. D. Rose, D.
A. Ritchie, A. K. Savchenko, and T. G. Griffiths, Phys. Rev. Lett. {\bf
80}, 1292 (1998).
\bibitem{zala} G. Zala, B. N. Narozhny, and I. L. Aleiner, Phys. Rev. B {\bf
64}, 214204 (2001); G. Zala, B. N. Narozhny, and I. L. Aleiner, Phys. Rev. B
{\bf 65}, 020201(R) (2002).
\bibitem{AbKrSa01}
For a review see E. Abrahams, S. V. Kravchenko, and M. P. Sarachik, Rev.
Mod. Phys. {\bf 73}, 251 (2001); S. V. Kravchenko and M. P. Sarachik,
Rep. Prog. Phys. {\bf 67}, 1 (2004).
\bibitem{Simonian98}
D. Simonian, S. V. Kravchenko, M. P. Sarachik, and V. M. Pudalov, Phys.
Rev. B {\bf 57}, R9420 (1998);
D. Simonian, S. V. Kravchenko, K. M. Mertes, M. P. Sarachik, and V. M.
Pudalov, Physica B {\bf 256-258}, 607 (1998).
\bibitem{Vitkalov01}
S. A. Vitkalov, H. Zheng, K. M. Mertes, and M. P. Sarachik, Phys. Rev.
Lett. {\bf 87}, 086401 (2001).
\bibitem{Shashkin01} A. A. Shashkin, S. V. Kravchenko, V.
T. Dolgopolov, and T. M. Klapwijk, Phys. Rev. Lett. {\bf 87}, 086801 (2001).
\bibitem{Shashkin02} A. A. Shashkin, S. V. Kravchenko, V.
T. Dolgopolov, and T. M. Klapwijk, Phys. Rev. {\bf B 66}, 073303 (2002).
\bibitem{Dolg92}
V. T. Dolgopolov, G. V. Kravchenko, A. A. Shashkin, and S. V. Kravchenko,
JEPT Lett. {\bf 55}, 733 (1992).
\bibitem{Simonian97}
D. Simonian, S. V. Kravchenko, M. P. Sarachik, and V. M. Pudalov, Phys.
Rev. Lett. {\bf 79}, 2304 (1997).
\bibitem{Pudalov97}
V. M. Pudalov, G. Brunthaler, A. Prinz, and G. Bauer, Pis'ma Zh. Eksp.
Teor Fiz. {\bf 65}, 887 (1997) [JEPT Lett. {\bf 65}, 932 (1997)].
\bibitem{Yoon99}
J. Yoon, C. C. Li, D. Shahar, D. C. Tsui, and M. Shayegan, Phys. Rev.
Lett. {\bf 84}, 4421 (2000).
\bibitem{okamoto} T. Okamoto, K. Hosaya, S. Kawaji, and A. Yagi, Phys. Rev.
Lett. {\bf 82}, 3875 (1999).
\bibitem{vitkalovSdH} S. A. Vitkalov, H. Zheng, K. M. Mertes, M. P. Sarachik,
and T. M. Klapwijk, Phys. Rev. Lett. {\bf 85}, 2164 (2000); S. A. Vitkalov,
M. P. Sarachik, and T. M. Klapwijk, Phys. Rev. 
{\bf B 64}, 073101 (2000).
\bibitem{AgKo98}
N. V. Agrinskaya and V. I. Kozub, Solid State Comm., {\bf 108}, 355
(1998).

\end{thebibliography}
\end{document}